\documentclass[aip, reprint,groupedaddress]{revtex4-1}
\usepackage[pdftex]{graphicx}
\usepackage[english]{babel}
\usepackage[T1]{fontenc}
\usepackage{amsmath}

\usepackage[colorlinks,linkcolor=blue,citecolor=blue,urlcolor=blue]{hyperref}
\hypersetup{pdfauthor={P.~A. Mu\~noz},
pdftitle={
Two-stage electron acceleration by 3D collisionless guide field magnetic reconnection
}}
\begin{document}

\title{
Two-stage electron acceleration by 3D collisionless guide field magnetic reconnection
}
\author{P.~A. Mu\~noz}
\email{munozp@mps.mpg.de}
\affiliation{Max-Planck-Institut f\"ur Sonnensystemforschung, D-37077 G\"ottingen, Germany}
\altaffiliation{Max-Planck/Princeton Center for Plasma Physics}

\author{J. B\"uchner}
\affiliation{Max-Planck-Institut f\"ur Sonnensystemforschung, D-37077 G\"ottingen, Germany}
\altaffiliation{Max-Planck/Princeton Center for Plasma Physics}

\begin{abstract}
	We report a newly found two-stage mechanism of electron acceleration
	near X-lines of 3D collisionless guide-field
	magnetic reconnection in the non-relativistic regime
	typical, e.g., for stellar coronae.
	We found that after electrons are first pre-accelerated
	during the linear growth of reconnection,
	they become additionally accelerated in the course of
	the nonlinear stage of 3D guide-field magnetic reconnection.
	This additional acceleration is due to the filamentation of electric and magnetic fields
	caused by streaming instabilities.
	In addition to enhanced parallel electric fields,
	the filamentation leads to additional curvature-driven electron acceleration
	in the guide-field direction.
	As a result, part of the the accelerated electron spectra becomes
	a power law with a spectral index of~$\sim-1.6$ near the X-line.
	This second stage of acceleration due to nonlinear reconnection
	is relevant for the production of energetic electrons in, e.g., thin current
	sheets of stellar coronae.
\end{abstract}

\maketitle

\section{Introduction\label{sec:intro}}
One of the unsolved puzzles in the Universe is the acceleration of electrons
to high energies in a wide variety of astrophysical objects.
They are remotely detected by the high-frequency electromagnetic radiation of,
e.g., hard X-rays from the solar chromosphere during
solar flares~\citep{Zharkova2011,Vilmer2012},
or directly observed by in-situ spacecrafts measurements
like by the ongoing MMS mission~\citep{Burch2016}.
The in-site measurements also provides the typical plasma and fields
conditions at the acceleration sites~\citep{Chen2016j}.
Common for those observations is the presence of current sheets (CSs)
and magnetic reconnection through them. Reconnection can, in principle,
accelerate electrons at the expense of the annihilation of
magnetic flux and energy~\citep{Yamada2010}.

Mechanisms of efficient acceleration of non-relativistic electrons
by guide-field magnetic reconnection
typical for, e.g., stellar coronae are, however, still not clear.
They were investigated, e.g.,
by test particle calculations using prescribed reconnection fields
usually obtained by MHD simulations~\citep{Zhou2015}.
Calculations using resistive MHD fields revealed effective
electron acceleration mainly by magnetic-field-aligned (parallel)
reconnection electric fields ($E_{\parallel}$) both near
single~\citep{Gordovskyy2010}
and multiple X-lines~\citep{Zhou2016d},
in stochastic CSs~\citep{Vlahos2004},
in turbulent fields~\citep{Kowal2012},
and also via a two-stage energization process~\citep{Dalena2014}.
Test particle calculations, however,
do not take into account the feedback
of the energized electrons to the plasma.
In addition, they usually overestimate the electron acceleration since the
parallel electric fields obtained by an ad-hoc assumed ``anomalous'' resistivity,
or numerical effects (dissipation),
are much larger than in collisionless astrophysical plasmas~\citep{Buchner2006,Kowal2012}.
To avoid exaggerated  parallel electric fields due to assumed ``anomalous'' resistivities,
or the usually high and
not well controlled numerical resistivity of MHD codes, self-consistent kinetic
investigations have to be carried out.
But those kinetic numerical simulations are, however,
usually limited to relatively small spatial scales.
Fully-kinetic PIC-code simulations of already relativistic electron-proton plasmas
revealed effective
electron acceleration by reconnection in 2D-~\citep{Melzani2014a,Guo2016a}
and 3D-~\citep{Totorica2016} configurations
as well as for relativistic pair
plasmas~\citep{Zenitani2001,Cerutti2014a,Sironi2014,Guo2015g}.
This way, power-law electron energy spectra were found in magnetically dominated relativistic plasmas,
where the energy available for acceleration is orders of magnitude larger than the plasma rest energy.
For non-relativistic electron-proton plasmas, however,
self-consistent kinetic investigations revealed so far only a weak electron energization by
the reconnection electric field near X-lines in 2D configurations,
which might be enhanced by mechanisms such as surfing and parallel electric fields
in the separatrices~\citep{Hoshino2001,Pritchett2006,Wan2008,Egedal2012}.
In contracting magnetic islands (plasmoids) of long CSs,
first order Fermi-type acceleration was detected
in 2D~\citep{Drake2006a,Dahlin2014,Li2015c,Dahlin2016a}
and also in 3D configurations~\citep{Dahlin2015,Dahlin2017}.
However, power-law energy spectra were not found.
Stochastic Fermi acceleration
due to the interaction of multiple magnetic islands was observed in
the solar wind~\citep{Khabarova2017b} and
found in  simulations of turbulent reconnecting plasmas~\citep{Hoshino2012a,Drake2013a}.
Fermi acceleration is, however, suppressed in strong guide-field (low plasma-$\beta$)
magnetic reconnection~\citep{Dahlin2016a,Wang2017t}.
On the other hand, the problem of non-relativistic electron acceleration in guide field reconnection
is critical to understand, e.g., the observed X-ray spectra in solar and other stellar coronae,
which require electron acceleration out of a thermal distribution.

We noted, however, that all the previous research did not take into account the
nonlinear evolution of 3D guide-field magnetic reconnection, which
in thin current sheets causes a filamentation in the guide-field direction.
In fact, we found that this filamentation in strong guide field
($B_g$, larger than the asymptotic, upstream, reconnection magnetic field $B_{\infty y}$)
reconnection  causes a so far unknown, second-stage electron acceleration, which
generates power-law electron spectra at single X-lines.

\section{Method}
We describe the nonlinear evolution of guide-field reconnection and the
consequent 3D structure formation by fully-kinetic, relativistic PIC-code
simulations using the code ACRONYM~\citep{Kilian2012}.
To allow periodic boundary conditions in all three directions,
we initialize two force-free equilibrium sheets of currents
flowing in opposite directions and sufficiently separated, avoiding
their interaction at short time-scales~\citep{Munoz2015}.
We focus on the investigation of single X-line reconnection
in one of the current sheets.
For this sake, we carried out 3D kinetic simulations with various parameters.
We illustrate our findings by simulations results obtained for
an ion-electron mass ratio  $m_i/m_e=100$,
equal electron and ion temperatures $T_i=T_e$,
a plasma beta $\beta_e=\beta_i = 2\mu_0 n_0 k_B T_i/B_T^2=0.016$,
a ratio of the electron thermal speed to the speed of light of $v_{th,e}/c=0.1$
and a relative guide field strength  $b_g=B_g/B_{\infty y}=2$.
$B_T=B_{\infty y}\sqrt{1+b_g^2}$ is the initially constant
total magnetic field
and the CS half-width is $L=0.25d_i$,
where $d_i=c/\omega_{pi}$ is the ion inertial length
and $\omega_{pi}$ is the ion plasma frequency.
The initial ion and electron number densities are constant
and equal ($n_i=n_e=n_0$).
All other parameters and quantities can be obtained from those, given above.
Absolute values can be deduced by choosing, as usual, the electron plasma
frequency or density for the plasmas of interest
(solar corona, solar wind, Earth's magnetosphere, etc).

The simulation box
$L_x\times L_y\times L_z$ covers a relatively small physical domain
$4\,d_i\times 8\,d_i\times 16\,d_i$.
The numerical mesh is spanned over $256\times 512\times 1024$ grid points.
We intentionally choose a simulation domain short enough
in the $y$-direction in order to avoid multiple magnetic islands formation and
contraction, and thus to investigate
the two-stage electron acceleration process at a single X-line.
The plasma is represented by
200 macro-particles per cell (PPC; 100 per specie),
which corresponds to a total of $2.7\cdot 10^{10}$ particles in the simulation box.
To verify the convergence of our results, we ran simulations with
smaller and larger number of PPC.
We found that for less than 25 PPC,
numerical collisions due to the PIC shot noise start
slow down the electrons numerically, causing heating
instead of electron acceleration (see also~\citet{May2014}).
Reconnection is triggered by a
3D perturbation of the magnetic field narrowly
localized  in the current direction ($z$) and with
a long (most unstable) tearing-Eigenmode wavelength in
the $y$-direction.

\section{Two-stage acceleration}

\begin{figure*}[!ht]
	\centering
		\includegraphics[width=0.99\linewidth]{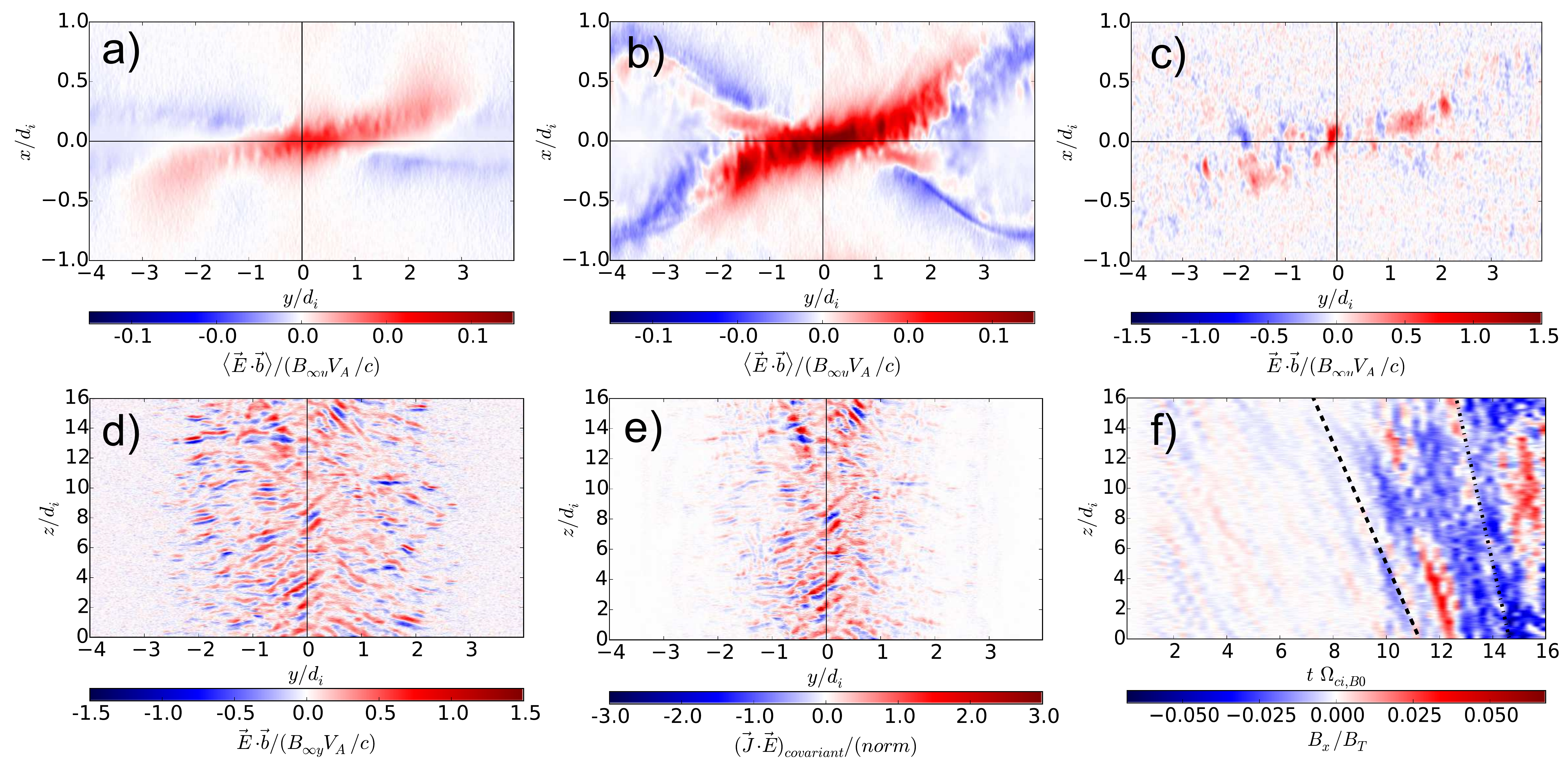}
	\caption{
	a) and b) Spatial distribution of $\langle E_{\parallel}\rangle$,
	the parallel electric field $E_{\parallel}(x,y)$ averaged over $z$
	at $t=10\,\Omega_{ci}^{-1}$ (a) and at $t=13.5\,\Omega_{ci}^{-1}$ (b)).
	c) $E_{\parallel}(x,y)$ at $t=13.5\,\Omega_{ci}^{-1}$ in the plane $z=2d_i$.
	d) $E_{\parallel}(y,z)$ in the central plane of the CS ($x=0$) at $t=13.5\,\Omega_{ci}^{-1}$.
	e) ($\vec{j}\cdot\vec{E})'(y,z)$
	in the CS central plane ($x=0$) at $t=13.5\,\Omega_{ci}^{-1}$,
	normalized to $en_0 v_{th,e}B_{\infty} V_A$.
	f) Time evolution of the magnetic perturbations $B_x$
	in the X-line  ($x=y=0$) along the guide-field direction $z$.
	\label{fig:structure_3d}}
\end{figure*}

In three-dimensions a local perturbation triggers a wave of
quasi-2D reconnection in the plane perpendicular to the guide field,
propagating in the guide-field direction~\citep{Buchner1999,Huba2002,Lapenta2006a,Jain2017b}.
Figs.~\ref{fig:structure_3d}a)-b) show the resulting
$\langle E_{\parallel}\rangle$, the parallel electric field $E_{\parallel}$
averaged over the $z$ direction in the plane $x-y$, at two characteristic times:
during the linear growth of reconnection ($t=10\,\Omega_{ci}^{-1}$) and in the middle of the nonlinear evolution  ($t=13.5\,\Omega_{ci}^{-1}$).
As usual, all PIC-quantities are
time-averaged over $0.1\,\Omega_{ci}^{-1}$ to remove
the high-frequency PIC shot noise.
Note the alignment of the average parallel electric field
along the low-density separatrix (see Fig.~\ref{fig:trajectories}),
where the density and also the current density $j_z$ are smaller
than the other separatrix due to the asymmetry
introduced by the guide field.
\begin{figure*}[!ht]
	\centering
		\includegraphics[width=0.99\linewidth]{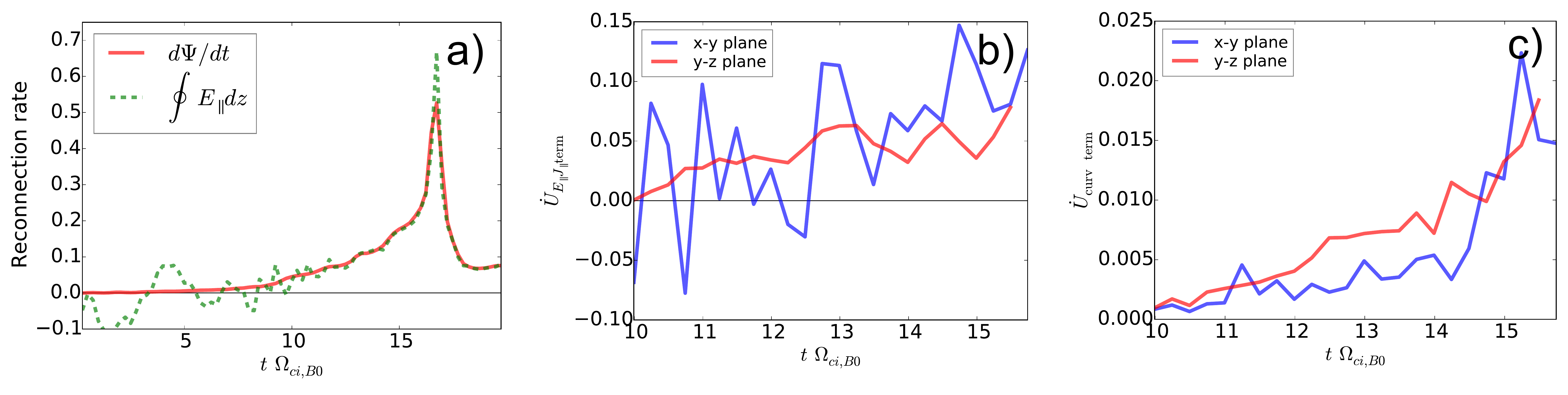}
	\caption{
	a) Normalized reconnection rates.
	b) Time evolution of the averaged (in two planes) parallel energization term in Eq.~\ref{eq:acceleration}.
	c) Time evolution of the averaged (in two planes) curvature energization term in Eq.~\ref{eq:acceleration}.
	The quantities in the $y-$ axis of b) and c) are normalized
	to $E_0J_0$, where $J_0=en_ev_{th,e}$.
	Note the different range in the $y$ axis.
	\label{fig:rec_rate_time_evolution}}
\end{figure*}

The maximum $\langle E_{\parallel}\rangle$
grows as the reconnection rate, becoming
$\langle E_{\parallel}\rangle=0.08E_0$ at  $t=10\,\Omega_{ci}^{-1}$
 and  $0.15E_0$
at $t=13.5\,\Omega_{ci}^{-1}$, respectively ($E_0=V_A B_{\infty y}$ and
$V_A$ is the Alfv\'en speed on $B_{\infty y}$).
Fig.~\ref{fig:rec_rate_time_evolution}a) shows the reconnection rate
(normalized to $E_0$),
calculated both as the rate of change of magnetic flux
between the X and O lines (red line) and independently calculated as
a path integral of the parallel electric field in a rectangle containing
the X- and O- lines (green dashed line).
Before $t\sim 13\,\Omega_{ci}^{-1}$, reconnection is at its linear growth stage.
$E_{\parallel}$ still exhibits a typical 2D structure.
It only mildly accelerates electrons and it does not
generate power-law spectrum of energetic electrons.
After $t\sim13.5\,\Omega_{ci}^{-1}$, however,
the reconnection rate is strongly enhanced during the nonlinear growth
of reconnection as a result of the CS thinning.
A peak value of the reconnection rate (above $0.6E_0$) is reached at $t\sim16.5\,\Omega_{ci}^{-1}$.
It is due to the depletion of the available magnetic flux because of the
periodic boundary conditions in the $x$ direction,
which later let the CSs to influence each other when
the magnetic island around the O-line starts to influence the
the X-line region of the first CS.
This saturation by technical limitations causes
the reconnection to drastically reduce because the available
magnetic flux is exhausted.
As a result, reconnection is a non-stationary transient process,
lasting as long as new flux is available, as it is typically observed by energetic particles.
A longer duration  or even steady-state reconnection  can be reached
only if more magnetic flux is provided,
but that is not an issue for the
short-time scale acceleration, described here.

During the nonlinear stage of reconnection, thinning CSs can
develop streaming and shear-flow-driven instabilities.
At their nonlinear stage, the resulting plasma waves cause a filamentation
of the electromagnetic fields along the guide-field direction $z$
(see \citet{Munoz2017} for details about the evolution of the instabilities,
turbulence and structure formation).
In the reconnection plane $x$--$y$, the filamentation of $E_{\parallel}$ appears
in patchy structures, different at each slice/plane along $z$. For example, Fig.~\ref{fig:structure_3d}c)
shows $E_{\parallel}(x,y)$ in the plane $z=2d_i$ at  $t=13.5\,\Omega_{ci}^{-1}$, while
Fig.~\ref{fig:structure_3d}d) shows
$E_{\parallel}(y,z)$ in the plane $x=0$  (the CS center) also at $t=13.5\,\Omega_{ci}^{-1}$.
The signs of $E_{\parallel}(y,z)$ alternate as indicated by red/blue colors.
The maximum absolute value of the parallel electric field is
$\vert E_{\parallel,max}\vert \sim 1.5E_0$.
This exceeds the average parallel electric field $\langle E_{\parallel}\rangle$
by a factor of ten.
Note that
similar filamentary structures were obtained before by other PIC-code simulations
near X-lines~\citep{Drake2003,Che2011,Wendel2013}
and near the separatrices of reconnection~\citep{Pritchett2012,Roytershteyn2012,Markidis2012a}.
Meanwhile they were observed in-situ in the Earth's
magnetosphere~\citep{Khotyaintsev2010,Mozer2010a,Ergun2016}.
The work done by the electromagnetic field on the particles is given by
$(\vec{j}\cdot\vec{E})'$.
The prime (') indicates that it is
calculated in the electron frame of reference and with
an extra compensating factor due to the charge separation (see \citet{Zenitani2011}).
This is significant in this regime where the plasma frequency $\omega_{pe}$ is of the same order as the electron cyclotron frequency $\Omega_{ce}$.
The quantity $(\vec{j}\cdot\vec{E})'$ is also spatially filamented, as one can
see in Fig.\ref{fig:structure_3d}e).
Positive values ($(\vec{j}\cdot\vec{E})'>0$) indicate local dissipation
(transfer of energy from the electromagnetic fields to particle heating), while
negative values ($(\vec{j}\cdot\vec{E})'<0$) mean
the opposite: a local transfer of energy from  particles to the electromagnetic field.
On (spatial) average, however, there is a net energy transfer from the electromagnetic fields to the particles
($\langle\vec{j}\cdot\vec{E}\rangle'>0$).

The time evolution of the filamentary structures
is illustrated by Fig.~\ref{fig:structure_3d}f).
The Figure shows a time series of line cuts of the
magnetic field perturbation $B_x(z)$ along $z$ for $x=y=0$, the center of the X-line,
stacked sequentially for each time.
The dashed and dotted lines correspond to the initial
(current-carrying) electron drift
speed ($V_{De}= 4\,V_A$) and twice that value ($8\,V_A$), respectively.
They demonstrate that the filamentary structures propagate at
the instantaneous electron drift speed.
The latter increases in the course of the CS thinning,
reaching a speed of $8\,V_A$  at $t=13.5\,\Omega_{ci}^{-1}$.
This spreading of localized reconnection was
previously investigated  by two-fluid
investigations~\citep{Shepherd2012}, in experiments~\citep{Dorfman2014},
in Hall-MHD~\citep{Huba2002}, EMHD-~\citep{Jain2017b}
and fully-kinetic CSs simulations~\citep{Buchner1999,Lapenta2006a}.
Reconnection spreads at the electron drift speed
($4\,V_A$, our case) or
at the speed of shear Alfv\'en waves  in the guide field ($2\,V_A$),
whatever is the fastest.

\begin{figure*}[!ht]
	\centering
		\includegraphics[width=0.99\linewidth]{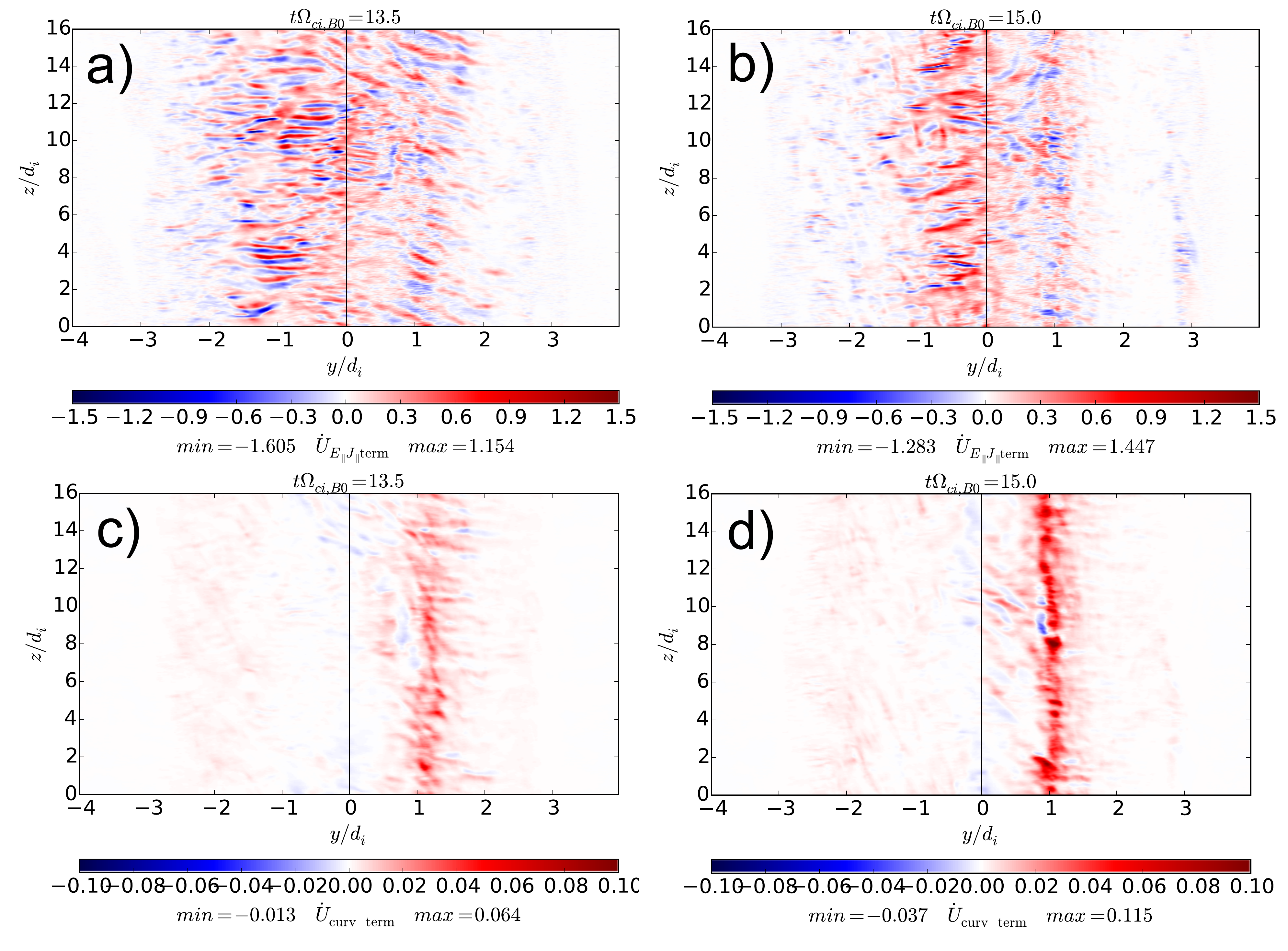}
	\caption{Electron energy change rate
	in the plane $y-z$ slightly off the center ($x=-0.1d_i$).
	Shown are the contributions of the
	first two terms of Eq.~\ref{eq:acceleration}.
	Top row: $E_{\parallel}J_{\parallel}$  due to parallel electric fields
	at a) $t=13.5\,\Omega_{ci}^{-1}$ and b) $t=15\,\Omega_{ci}^{-1}$.
	Bottom row: Energy change by curvature-driven acceleration at
	c) $t=13.5\,\Omega_{ci}^{-1}$ and d) $t=15\,\Omega_{ci}^{-1}$.
	See normalization in the text.
	\label{fig:acceleration_terms}}
\end{figure*}

As long as the electrons are fully magnetized, their motion can be described
in the guiding-center approximation. This is the case
when the electron gyroradius stays smaller than the typical
spatial scales of the magnetic field variation.
In the guide-field reconnection case, considered here,
the maximum electron gyroradius always satisfies this condition:
the $\kappa$ parameter~\citep{Buchner1989,Buchner1991}, defined as the ratio of
the curvature radius of magnetic field lines and electron gyroradius,
stays always above $10$. A more detailed discussion about the validity
of this approximation is given in Appendix~\ref{sec:appendix3}.

Changes of
the electron energy can be quantified in
the guiding-center approximation as~\citep{Northrop1963,Dahlin2014}:
\begin{align}\label{eq:acceleration}
\frac{dU}{dt} & = E_{\parallel}J_{\parallel}
+\left(p_{e,\parallel}+ m_e n_e u_{e,\parallel}^2\right)\vec{u}_{\vec{E}}\cdot\vec{\kappa}\\
&+\frac{p_{e,\perp}}{B}\left(\frac{\partial B}{\partial t} + \vec{u}_{\vec{E}}\cdot\vec{\nabla}B\right). \nonumber
\end{align}
Here $U$ is the electron kinetic energy density,
$\vec{u}_{\vec{E}}$ is the $\vec{E}\times \vec{B}$ drift speed,
$p_{e,\parallel}$/$p_{e,\perp}$  are the parallel/perpendicular
components of the electron pressure,
$u_{e,\parallel}$ is the parallel electron bulk flow velocity
and $\vec{\kappa}=(\vec{B}/B)\cdot\vec{\nabla}(\vec{B}/B)$
is the curvature of the magnetic field lines, a vector different from the
aforementioned curvature parameter $\kappa$.
The first term on the right-hand side (r.h.s.) of Eq.~\eqref{eq:acceleration}
describes the effect of parallel electric fields,
the second is due to the magnetic field curvature $\vec{\kappa}$.
The third term arises due to magnetic field inhomogeneities.

\begin{figure*}[!htbp]
		\centering
		\includegraphics[width=0.99\linewidth]{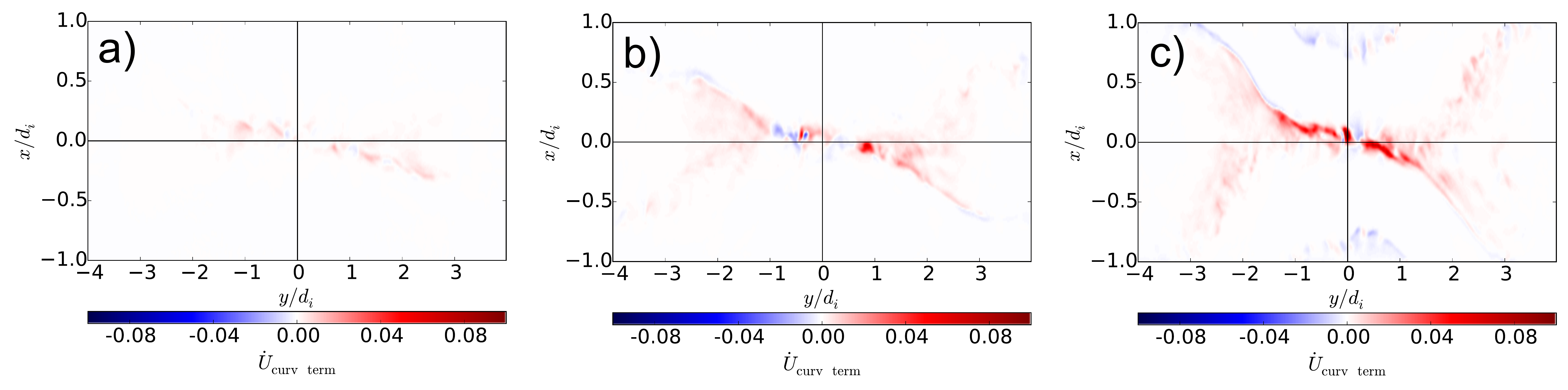}
	\caption{Electron energy change rate due to the
	curvature term in Eq.~\ref{eq:acceleration} at different times.
	a) $t\Omega_{ci}=11$. b) $t\Omega_{ci}=13$.  c) $t\Omega_{ci}=15$.
	The plots are at the $x-y$ plane at the slice $z=2d_i$.
	Same normalization as Fig.~\ref{fig:acceleration_terms}.
	\label{fig:fermi_term_xy_plane}}
\end{figure*}

Figs.~\ref{fig:acceleration_terms} shows the spatial distribution of
the main contributions to the acceleration of all electrons (top and bottom rows)
in a plane $y-z$, slightly off the CS midplane ($x=-0.1d_i$), at two different times.
In this Figure, all the quantities are normalized to $E_0J_0$, where $E_0=B_0V_A$ and
$J_0=en_ev_{th,e}$.
The contribution of the third term in the r.h.s. of Eq.~\eqref{eq:acceleration}
is not shown in the Figure because it is negligible, staying always an order of magnitude smaller compared to
the other two terms.
Panels a)-b) and c)-d) in Fig.~\ref{fig:acceleration_terms}
clearly show that the parallel electric field term can locally energizes
electrons more efficiently than the curvature-driven acceleration.
But it can also decelerate particles, while the
curvature acceleration almost always energizes the electrons.
The maximum normalized energy gain due to the
curvature term is $\dot{U}_{\rm curv,term}\sim 0.06$ at  $t=13.5\,\Omega_{ci}^{-1}$,
increasing up to  $\dot{U}_{\rm curv,term}\sim 0.11$ at $t=15\,\Omega_{ci}^{-1}$,
reached near $y\sim 1d_i$ and always with positive values.
Fig.~\ref{fig:fermi_term_xy_plane} illustrates (in the $x$--$y$ plane) that
off the CS center and near the high-density separatrix,
curvature-driven acceleration is stronger than at the very X-line.
This is in contrast to the parallel
electric field acceleration, which decreases
away from the X-line and it is stronger in the low-density
rather in the high-density separatrix (see Figs.~\ref{fig:structure_3d}b)-c)).

Another comparison of parallel-electric-field and curvature-driven acceleration
is shown in Figs.~\ref{fig:rec_rate_time_evolution}b)-c).
The Figures display the evolution of two averages of the main contributions to the acceleration with time.
The blue line is obtained by calculating the acceleration in a rectangular region
in the plane $x$--$y$ (at a particular $z$-slice), containing
the bottom left high-density separatrix (see Fig.~\ref{fig:fermi_term_xy_plane},
$\Delta x=0.2d_i$ below the CS midplane and between $y=[0,1.5]d_i$).
The red line in Fig.~\ref{fig:rec_rate_time_evolution}c) is calculated in a rectangle
located in the plane $y$--$z$ between $y=[0.5,1.5]d_i$ and all along $z$ for
the same $x$-slice used in Figs.~\ref{fig:acceleration_terms}c)-d)).
The two averages illustrate that the electrons
are not accelerated homogeneously all over, but in the filamentary structures.
There, the curvature-driven  acceleration is the strongest, in particular in the
dynamically changing region of the high-density separatrix close to the X-line (see
also some typical particle trajectories in Figs.~\ref{fig:trajectories}-\ref{fig:trajectories_new}).
Thus, the blue line in Fig.~\ref{fig:rec_rate_time_evolution}c) shows that the
area in which curvature-driven energization takes place is strongly enhanced
after $t\sim14\Omega_{ci}^{-1}$ (the average jumps by a factor of 4). The red line,
on the other hand, demonstrates that the magnitude of its contribution to the
energization increases steadily until $t\sim(13.5-14)\Omega_{ci}^{-1}$, becoming
enhanced by a factor of two afterwards.
A similar diagnostic for the parallel acceleration is shown by
the blue line in Fig.~\ref{fig:rec_rate_time_evolution}b). It indicates the
highly oscillatory nature of the parallel acceleration in the plane $x$--$y$,
with alternating signs before $t\sim13\Omega_{ci}^{-1}$.
It slightly increases later,
but with maximum values of the same order as the amplitude of the oscillations
around the average.
The red line in Fig.~\ref{fig:rec_rate_time_evolution}b) shows a steady increase up to
$t\sim13\Omega_{ci}^{-1}$ and later some oscillations around a given mean value.
The mean value reaches about three times the maximum average obtained by
curvature-driven acceleration.
Diagnostics for other planes $x$--$y$ or $y$--$z$ reveal a
similar behavior of curvature-driven or parallel-electric-field contributions to
the electron acceleration, just at slightly different levels.
This comparison indicates that the parallel electric field contribution
to the electron energization is still dominant at the nonlinear stage of reconnection,
but the curvature-driven acceleration
starts playing an important role.
It is better correlated with the localized particle acceleration in the
high-density separatrix,
contributing, thus, to the energization.
The increase of the curvature-driven acceleration coincides
with the most efficient electron acceleration after
$t\sim13.5\Omega_{ci}^{-1}$ (see Figs.~\ref{fig:trajectories}-\ref{fig:trajectories_new}).
On the contrary, the average of the parallel electric field contribution is not related
(it is stronger at the X-line and oriented towards the low-density separatrix) and
it does not change significantly at the nonlinear stage of reconnection.

Note that the nonlinear filamentation process is essential for the efficiency of the curvature-driven acceleration,
the second stage of acceleration.
This type of acceleration is efficient in sufficiently thin current
sheets, in which thermal electrons are preaccelerated by parallel electric fields during the first stage of acceleration.
In thicker CSs, reconnection
stays laminar and filamentation does not take place.
In them, the curvature driven-acceleration, even though it
contributes to a positive net electron acceleration,
is practically negligible (for details, see Appendix~\ref{sec:appendix}).
To demonstrate this, we carried out a numerical experiment using the same simulation (with the same CS thickness), but stopping the evolution of electromagnetic fields before the filamentary structures
are formed (see Appendix~\ref{sec:appendix2}). This approach is equivalent to a test-particle method,
without the particle feedback to the electromagnetic fields.
It demonstrates that, in this way,
heated distributions and beam-like structures are formed but
without developing clear power-laws.
Therefore, only a self-consistent consideration of the
electromagnetic field feedback is necessary
for the two-stage acceleration.

\section{Electron trajectories and energy spectra}
\begin{figure*}[!ht]
	\centering
		\includegraphics[width=0.99\linewidth]{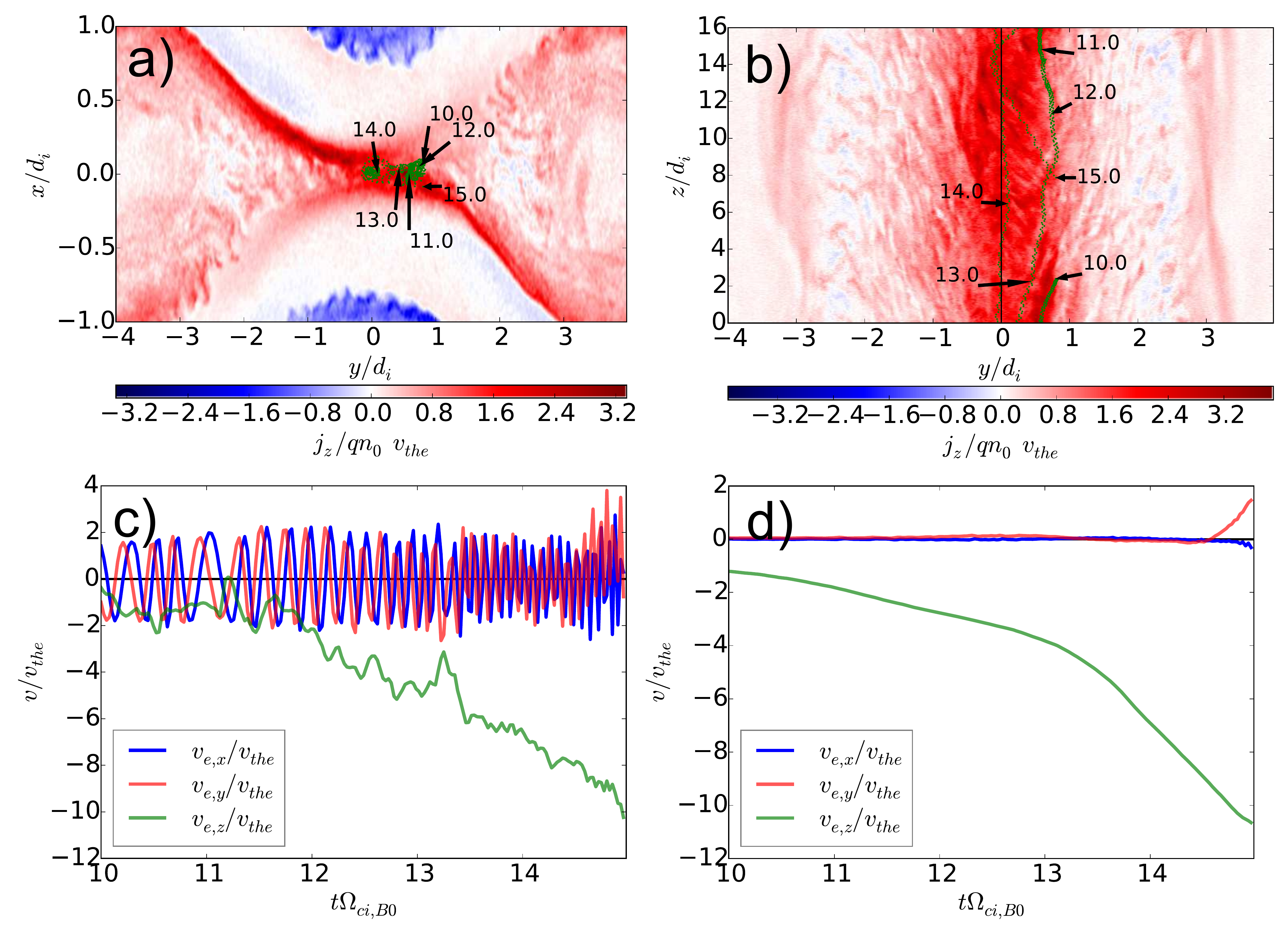}
	\caption{
	Projections of a typical trajectory  of a strongly energetic electron (green lines) to the
	a) plane $z=L_z/2$ and
	b) CS central plane $x=0$.
	Subsequent electron positions are indicated by time marks
	in units of $\Omega_{ci}^{-1}$.
	The colors depicts
	the current density $j_z$ at $t_f=15\,\Omega_{ci}^{-1}$.
	c) Evolution of the (four-)velocity components of the same electron.
	d) Time history of the ensemble-average (four-)velocity components of the
	$10^4$ most energetic electrons.
	\label{fig:trajectories}}
\end{figure*}

\begin{figure*}[!htbp]
	\centering
	\includegraphics[width=0.99\linewidth]{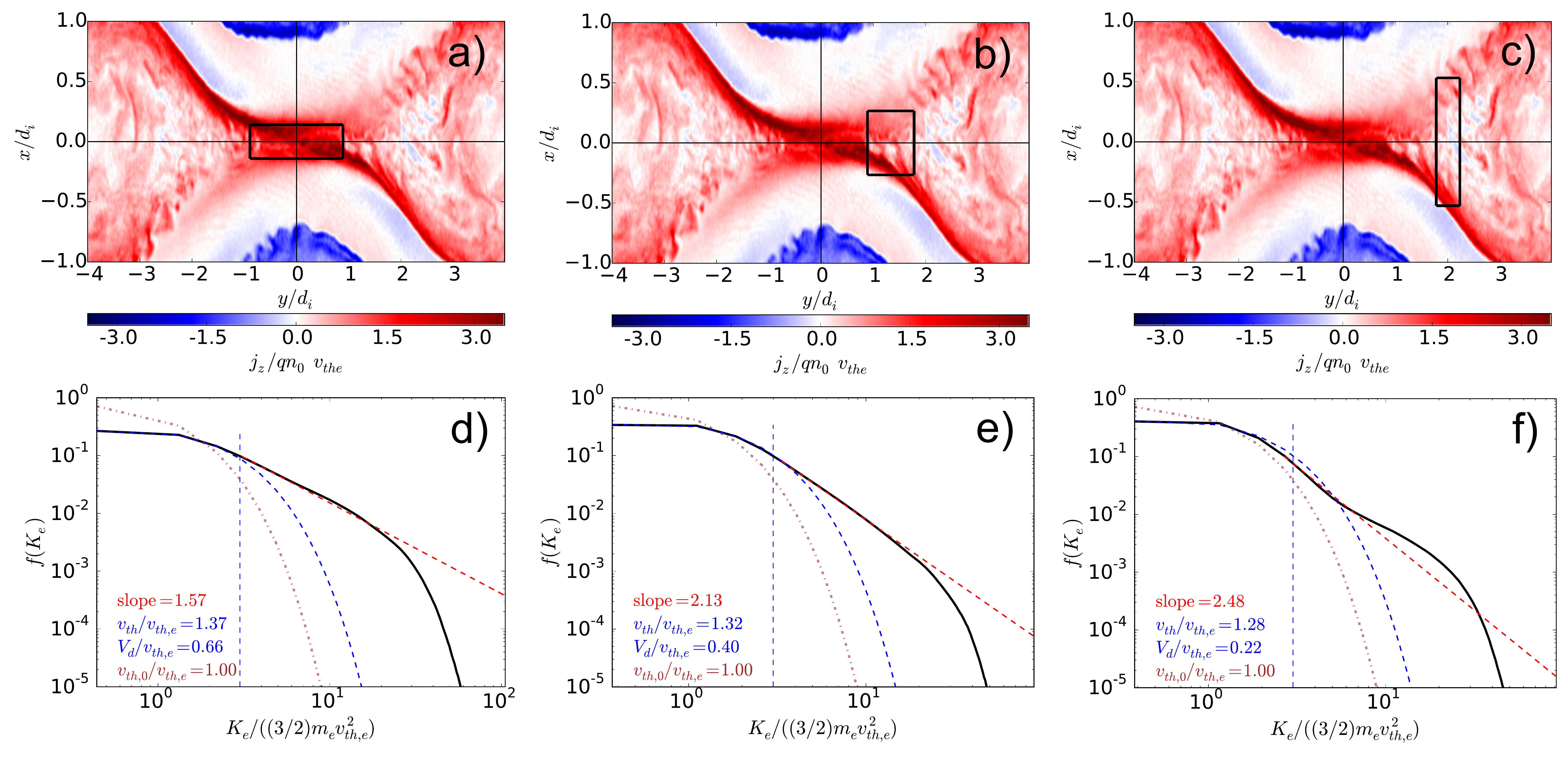}
	\caption{Electron energy spectra at $t_f=15\,\Omega_{ci}^{-1}$ (bottom row)
	obtained at different locations -- the black boxes in the top rows.
	Plots (a)-(c) show the current density $J_z$ in the plane $x-y$ at $z=L_z/2$ and for the time $t=t_f$.
	Brown dashed-dotted line: initial thermal distribution,
	blue dashed line: Maxwellian fit to the distribution, and red dashed line:
	power law fit to the energetic tail.
	\label{fig:spectra_locations}
	}
\end{figure*}

In order to demonstrate the mechanism of the two-stage
acceleration, Figs.~\ref{fig:trajectories}a)-b) depict projections of
a typical trajectory of one
self-consistently calculated by the PIC code strongly accelerated electron.
Fig.~\ref{fig:trajectories}c) shows the temporal evolution of its (four-)velocity
components.
Red and blue lines indicate
gyration in the magnetic  field,
while the green line shows the different efficiencies of the
electron acceleration during the two stages:
a mild pre-acceleration until
$t\sim (12-13.5)\,\Omega_{ci}^{-1}$, and the second acceleration stage
after the onset of the filamentation.
For every individual electron
the additional acceleration starts earlier or later,
depending on the motion phase of the electrons at the moment in which they enter the filaments.
Note that only pre-accelerated electrons can participate in the second-stage
energization due to the filamentation during the nonlinear evolution of reconnection.
The described two-stage acceleration mechanism
is sequential: only electrons already previously accelerated can
participate in the additional curvature-driven energization after
the filamentary structures starts to develop during the nonlinear stage
of reconnection.
Fig.~\ref{fig:trajectories}d) illustrate this by showing the time evolution
of the average velocities
components of the $10^4$ most energetic electrons diagnosed in
a small domain near the X-line (depicted in Fig.~\ref{fig:spectra_locations}a)) at $t\sim 15.0\,\Omega_{ci}^{-1}$.
It shows that the strongest acceleration takes place during
the nonlinear stage of guide-field reconnection, after filamentation has started ($t\sim 13.5\,\Omega_{ci}^{-1}$).
Out of the initially non-relativistic thermal distribution,
the fastest electrons now reach mildly relativistic
energies: the maximum four-velocity component $v_{e,z,max}$
corresponds to a particle speed of $0.74c$, i.e., to a
relativistic Lorentz factor of $\gamma\sim1.49$.
This corresponds to an increase of the  electron kinetic
energy by two orders of magnitude.

The electron energization can be characterized
by an effective electric field ($E_{\rm eff} =
(m_e /e) a_{\rm eff}$) which would cause the observed increase
$a_{\rm eff}= d \langle v_z \rangle / dt $ of the average
velocity $\langle v_z \rangle$ of these $10^4$ fastest  electrons.
In the pre-acceleration phase
(up to $t\sim 13.5\,\Omega_{ci}^{-1}$, see Fig.~\ref{fig:trajectories}d)),
a linear fit of $\langle v_z \rangle$ reveals $E_{\rm eff}\sim 2.0 \, E_0$.
At the nonlinear stage of reconnection, however, $E_{\rm eff}$ becomes
as large as $8.2\, E_0$.
This quantity exceeds several times even
the maximum value of $E_{\parallel}$ near the X-line (about $1.5\,E_0$).

\begin{figure*}[!ht]
		\centering
			\includegraphics[width=0.99\linewidth]{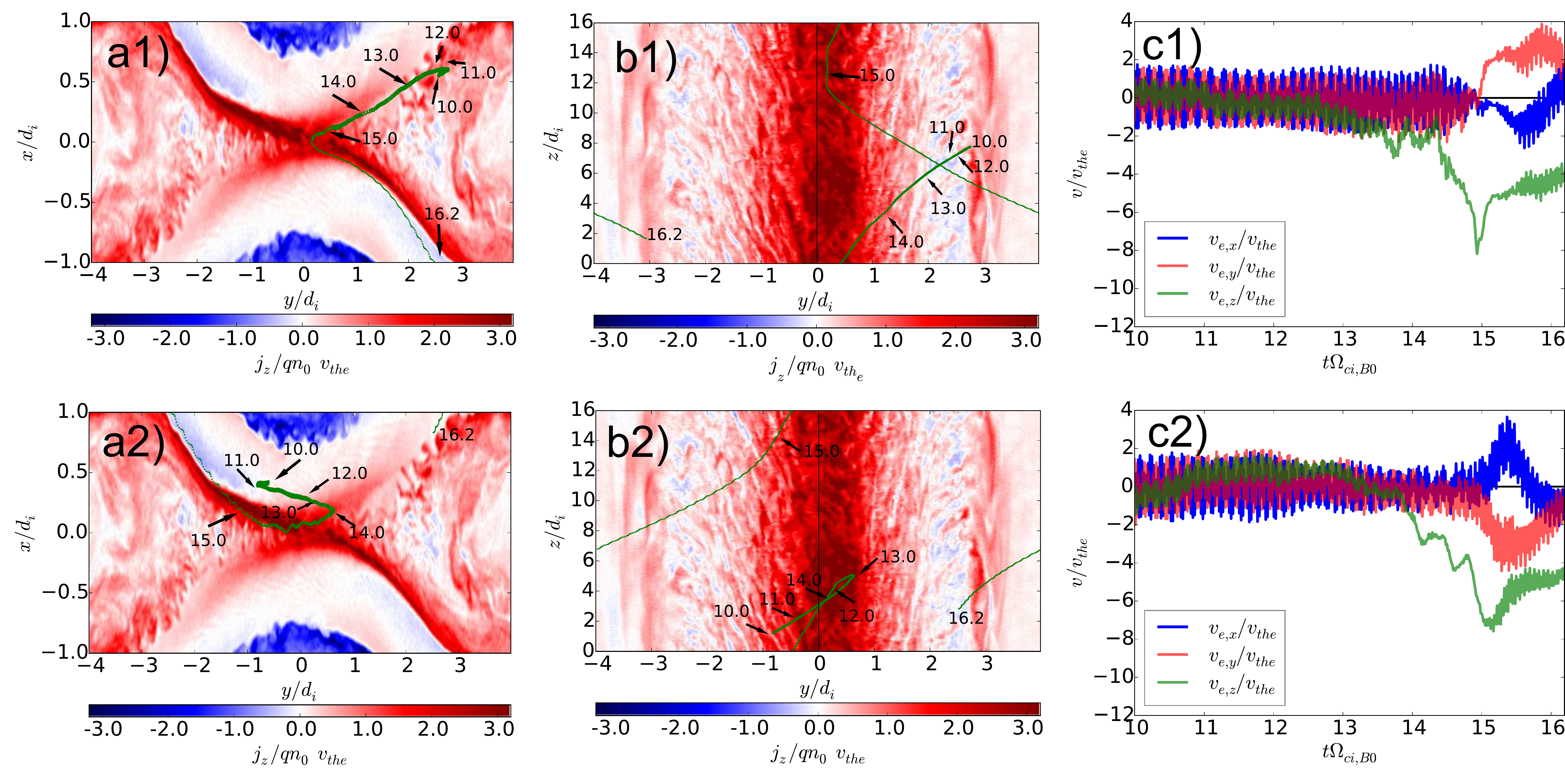}
		\caption{
		Each row depicts trajectories and velocities of two energetic
		electrons in a similar format as in Fig.~\ref{fig:trajectories}.
		Projections of trajectories
		(green lines)
		to the $x$--$y$ plane at $z=L_z/2$ (a1)-a2)) and to the
		$y$--$z$ plane at the CS midplane $x=0$ (b1)-b2)).
		c1)-c2): Time evolution of the (four-)velocity components of the
		electrons trajectories depicted in the panels a) and b)
		\label{fig:trajectories_new}}
	\end{figure*}

Note that the periodic boundary conditions in the guide field direction $z$, used in the simulations,
do not significantly affect the acceleration of even the most energetic electrons:
those with energies beyond the high energy end of the power-law
part of the spectrum cross the $z$-boundaries of the simulation domain
no more than
two or three times within the considered acceleration time.
Most of the electrons accelerated into the power law part of the spectrum,
however, do not even  cross the $z$--boundaries at all.
This can be seen in Fig.~\ref{fig:trajectories_new}, which shows
the trajectories and velocity components of two other
electrons accelerated to higher energies in a similar format to Fig.~\ref{fig:trajectories}.
The electron in the top row (Figs.~\ref{fig:trajectories_new}a1-b1-c1)) approaches
the X-line through the low-density separatrix (top right quadrant in the plot).
It becomes efficiently accelerated
in the $z$ direction only after it interacts with the fully developed filaments
This provides enhanced net curvature-driven acceleration mostly along the high-density separatrix
(bottom left quadrant in the plot).
After $t\sim 15\,\Omega_{ci}^{-1}$, the particle escapes from the X-line towards
the high-density separatrix region, converting part of the kinetic energy in the
$v_z$ component to the $v_x$ and $v_y$ components.
The electron in the bottom row of Figs.~\ref{fig:trajectories_new}a2-b2-c2)) starts
close enough to the X-line, approaching the low-density separatrix.
It then reverses its direction toward the other high-density separatrix
(top left quadrant in the plot), where it is additionally accelerated.
Each jump of the velocity component $v_z$
corresponds to the electron entering the filaments.
Note that both electrons do not cross the whole simulation box more than once
during its period of maximum acceleration (up to $t\sim 15\,\Omega_{ci}^{-1}$).

Fig.~\ref{fig:spectra_locations}d) shows the electron spectrum
at the saturated nonlinear stage  of reconnection ($t=t_f$),
obtained for a region (box) near the X-line (displayed in Fig.~\ref{fig:spectra_locations}a))
extending $0.25 d_i$ along $x$, $1.8 d_i$ along $y$, and
the full length in $z$ ($16d_i$).
The Maxwellian fit of the thermal
part of the spectrum (blue dashed line) indicates an electron heating by up to
$40\%$ above the initial temperature.
The non-thermal tail of the electron distribution is best fitted
by a power-law $f(K_e)=K_e^{-\alpha}$,
with the electron kinetic energy $K_e=m_ec^2(\gamma-1)$.
Simulations with more than 25 PPC reveal a stable spectral index
$\alpha\sim 1.6$, independent on any further increase on the number of particles per cell.
The power-law part of the spectrum (red dashed line) ranges over more than an order of
magnitude above the initial thermal energy.
Note that there is an exponential
cutoff at higher energies as predicted (solving simplified equations for the particle trajectories)
and also observed in simulations of relativistic pair plasmas~\citep{Bessho2012,Werner2015a}.

\begin{figure}[!ht]
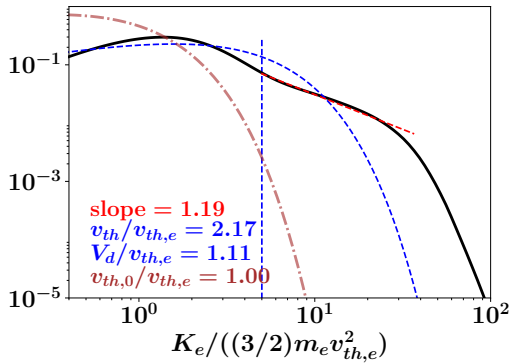

\centering
	\includegraphics[width=0.8\linewidth]{{{fig8}}}
	\caption{Electron energy spectra at $t_f=15\Omega_{ci}^{-1}$
	obtained by averaging over the whole simulation domain,
	including X-line, separatrices and reconnection exhaust.
	The format is the same as in Fig.~\ref{fig:spectra_locations}d-e-f).
\label{fig:energy_spectra_full}}
\end{figure}

As discussed previously for Fig.~\ref{fig:trajectories_new},
away from the X-line,
the fast electrons are decelerated and thermalized in the
exhaust regions of reconnection.
In them, the local electron energy spectra becomes thermalized
with the distance
from the X-line, the power-laws parts become steeper (softer
energy spectra).
Fig.~\ref{fig:spectra_locations}e) demonstrates this result by showing
the spectra obtained for electrons in a region $0.5 d_i$ along $x$ and $0.9 d_i$ along $y$ displayed
in Fig.~\ref{fig:spectra_locations}b).
The chosen box is twice as large along $x$ and half
sized along $y$ compared to the box chosen in Fig.~\ref{fig:spectra_locations}a),
in order to not only consider approximately the same number of particles but to capture the separatrices as well.
The resulting (off-center) spectrum (Fig.~\ref{fig:spectra_locations}e)
exhibits a steeper power-law section with an index of $-2.1$,
indicating less energetic, already thermalized electrons.
Further away from the X-line, the electron distribution
is even more thermalized (Fig.~\ref{fig:spectra_locations}f)),
while the power-law section is very short and steeper than
those observed closer to the X-line.
The latter electron distribution was obtained in the
region shown in  Fig.~\ref{fig:spectra_locations}c)
for a doubled box size along $x$ and halved size along $y$ compared to
the boxes used for Fig.~\ref{fig:spectra_locations}b).
We point out that the electron energy spectra
depends on the distance from the X-line.
This effect was probably overlooked
in previous investigations which, due to the
smaller number of particles used
were not able to diagnose the dependence of the electron spectra
on the distance from the X-line.
Spectra as they were obtained by averaging
over the whole simulation boxes do not show the
formation of a significant power-law spectra.
We demonstrate this by showing a spectra obtained
by averaging over the whole simulation box.
As one can see in Fig.~\ref{fig:energy_spectra_full},
such averaging hides the electrons energization by the two-stage acceleration, overlaying a large number of thermal electrons decelerated and thermalized away from the X-line.
There is mainly a heated component at low energies left after
such averages, while for higher energies only a very short power-law
is seen.

Note that the spectra shown in Fig.~\ref{fig:energy_spectra_full} do not
allow to immediately distinguish easily which part of the spectra is due to
the acceleration processes at the X-point.
In particular because the mechanisms leading to a slow down of particles
away from the X-line are usually exaggerated by using too
small simulation domains.
The periodicity along the $y-$ boundaries causes
the particles spending most of the time
close to the boundaries.
There, the counter-streaming plasma flows of the reconnection exhaust interact,
 creating turbulence.
This slows down energetic electrons in the simulation, but not in reality.
This is another reason, why averaging over the full simulation domains is not appropiate
for comparison with observations.
Previous simulations sometimes use extended domains
mainly along the $y-$ direction, in order to study Fermi acceleration
in contracting magnetic islands (see, e.g., Ref.~\cite{Dahlin2014}).
This way only a portion of the total particle number is affected by periodicity.
On the other hand, simulations with open boundary conditions along
the $y-$ direction can avoid that issue, but they are still uncommon.

\section{Conclusions and Discussion}
We have shown that collisionless 3D  guide-field
magnetic reconnection in a non-relativistic electron-proton plasma
accelerates electrons very efficiently via a two-stage mechanism
even at single X-lines, and even without contracting magnetic islands as
discussed, e.g., by  \citet{Drake2006a,Dahlin2014,Li2015c,Dahlin2016a,Dahlin2015,Dahlin2017}.
The first stage of pre-acceleration is due to the parallel electric fields
rising during the linear growth phase of reconnection.
It mildly pre-heats the electrons.
The second stage takes place during the nonlinear phase of reconnection,
which is characterized by a filamentation of the thinning current sheet.
At this stage the acceleration is enhanced by
curvature-driven acceleration.
The electron energy spectrum is characterized
by electron heating and the rise of a power-law with a spectral index
of up to $\alpha\sim-1.6$ near the X-line.
Such power-law spectra were meanwhile observed at reconnection sites by
{\it in-situ} observations in the Earth's magnetosphere~\citep{Turner2016}.
In the solar corona the electron spectra, deduced
from hard X-ray emission of flares, also indicate such power law
electron energy spectra~\citep{Holman2003,Kontar2014}.
While our findings do not explain the heating of the solar coronal plasma
to temperatures of millions of $K$.
Instead, they show that in addition to heating thin current sheets,
reconnection causes power-law electron spectra
in stellar coronae and other astrophysical environments.
Those electrons might further act, e.g., as seed particles required for Fermi-type
acceleration processes in collisionless shocks.

Note that some features of the second acceleration stage resemble
aspects of first~\citep{Fermi1949} and  second order~\citep{Davis1956,Parker1958b}
 Fermi acceleration.
It is, however, neither characterized by bulk plasma flows nor
by a stochastic electron motion.
Instead it is due to the filamentation of electromagnetic fields
by reconnection in thin current sheets, with contributions of curvature-driven
and parallel $E$-field acceleration, while Fermi acceleration usually depends
on elastic particles bounces~\citep{Balogh2013,Pisokas2017},
different from the dissipative processes present in filamented current sheets.

The extent of the power-law section of the electron spectra caused by the two-stage acceleration extends  with the length of the X-line.
As usual, the high energy cutoff of a power-law distribution
depends on the loss mechanism.
For acceleration in current sheets, this is mainly the electron escape time from the reconnection region.
After the electrons escape from the X-line region, they are thermalized
along the separatrices and in the exhaust regions of reconnection.
While the pitch-angle scattering is weak in strong guide
fields, the resulting thermalization steepens the energy spectra
thermalizing the distribution with the distance from the X-line.
This is the reason why electron energy spectra
obtained by averaging over large domains do not reveal the
power-law spectra by mixing the two-stage accelerated electrons with
electrons thermalized away from the X-line.

The two-stage acceleration by reconnection
is due to the filamentation of thin current sheets in a
low-beta plasma, which
is due to streaming and shear flow instabilities.
Such streaming instabilities, in particular the Buneman
instability, take place only in thin current sheets.
While the relative electron-ion streaming in them is initially inversely
proportional to the current sheet thickness,
in the course of the nonlinear evolution of reconnection,
the relative streaming starts to
significantly exceed the electron thermal speed
for the relatively cold electrons in a low beta-plasma.

Previous 3D PIC simulation studies of current sheet filamentation,
did not analyze the resulting electron acceleration (e.g., \citet{Che2011}).
Other simulations of similar filamentations reported strong electric fields
in the separatrices (see, e.g., \cite{Pritchett2013a}).
However, there the two-stage acceleration does not properly work, since
the electric field is too weak or vanishes completely.
Hence, the first stage of electron pre-acceleration is missing.

Other investigations of electron acceleration by reconnection
usually analyze ion-skin-depth (thick)
current sheets (e.g., \citet{Daughton2011, Pritchett2013a}).
But these current sheets usually do not thin down to scales at which filamentation take place.
We verified this point by investigating
thicker current sheets (see Appendix.~\ref{sec:appendix})
and we showed that such current sheets do not filament
and do not form power-law energy spectra.
We also showed that without feedback of the particles to the plasma,
no filamentation and curvature energization take place (see Appendix.~\ref{sec:appendix}).

It is well known that, although
parallel $E$-field acceleration takes place near X-lines,
it is not efficient in producing power-law spectra per se.
Plasmoid reconnection studies did not find filamentation along the
X-lines, but direct efficient acceleration by Fermi-type acceleration in
contracting magnetic islands/flux ropes  (see, e.g., \cite{Dahlin2015,Dahlin2017}).
A comparison of acceleration in plasmoids
and the two-stage acceleration in filaments can be addressed
only by studies using much larger simulation boxes.
The two-stage mechanism is efficient in
thin electron current sheets.
This can be shown only by simulations using a sufficient scale separation
of electrons and ions, with large simulation boxes
in the direction of the reconnection plane and in the
out-of-plane directions,
which makes them computationally very expensive.
In sufficiently large systems, with well-resolved
thin current sheets, the two-stage acceleration process efficiently
accelerates the electrons near the reconnection X-line.
As soon as the electrons move towards
the separatrices  or into the exhaust, they become decelerated and thermalized.
After entering magnetic islands,
they can be accelerated by Fermi-type processes in contracting islands.
As a result, most energetic electrons are found near X-lines and inside plasmoids,
accelerated by different mechanisms.
The relative efficiency of the two mechanisms
should depend on the macroscopic plasma parameters.
Fully-kinetic PIC-codes simulate relatively small spatial domains compared
to MHD-fluid treatments.
Nevertheless, with the box sizes and number
of particles used, we were able to fully reproduce the two-stage electron
acceleration and spectrum formation by guide-field
magnetic reconnection through thin current sheets.
Further changes of the box size of our PIC code-simulations did not practically  affect
the results anymore, the described acceleration is local compared
to the system size.
The domain size has only an indirect influence on the acceleration since it determines the duration of the reconnection process limited by the amount
of magnetic flux available.
But the two-stage acceleration takes place, anyways, at short time scales.
Therefore,
a longer duration of the reconnection process does not change
the principal conclusions about the efficiency of the two-stage electron acceleration
and power-law formation, only the maximum energy obtained  by
the accelerated particles.
In very large domains and over longer lasting reconnection processes,
electrons could be further accelerated, extending the range of the power-law
section of the spectrum to higher energies, but not changing the power-law index.
This corresponds to the well-known results about particle acceleration also by
collisionless shocks.

Further, note that PIC-code simulations results of non-relativistic phenomena
are controlled by dimensionless
parameter ratios rather than by the absolute values of the physical
quantities~\citep{Ryutov2012}.
Absolute values
are, therefore, of no fundamental importance but can rather be chosen
to match the particular physical scenario of interest by applying
the appropriate scalings.

\begin{acknowledgments}
	We acknowledge the developers of the ACRONYM code (Verein zur F\"orderung kinetischer Plasmasimulationen e.V.).
	In particular, we are most grateful to Patrick Kilian for his helpful discussions and valuable comments.
	We further acknowledge the Max-Planck-Princeton Center for Plasma Physics and the DFG Priority Program ``Planetary Magnetism'' SPP 1488 for funding.
	Computational resources were kindly provided by the PRACE project prj.1602-008
	in the Beskow cluster at the PDC/KTH, Sweden.
	We also used the Hydra cluster of the Max Planck Computing and Data Facility
	(MPCDF, formerly known as RZG) at Garching, Germany.
	We also thank the referees for their comments and suggestions
	that allowed us to make our investigation
	results better verified and proven.
\end{acknowledgments}

\appendix

\section{Influence of the current sheet thickness}\label{sec:appendix}

\begin{figure*}[!htbp]
	\centering
	\includegraphics[width=0.99\linewidth]{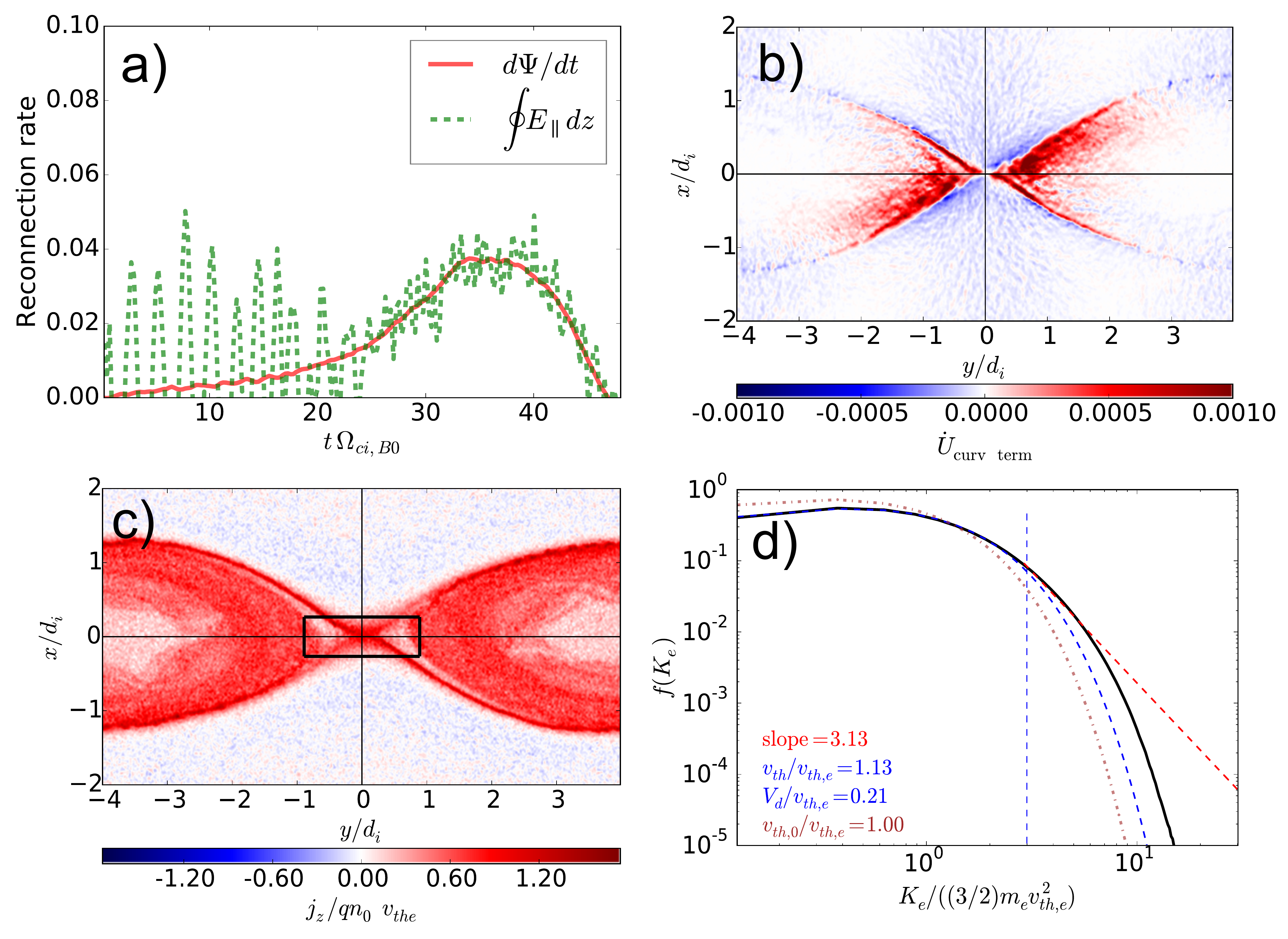}
	\caption{
	Results for a twice as thick current sheet compared to the one
	analyzed in the main text of the paper.
	a) Normalized reconnection rates, same legend as Fig.~\ref{fig:rec_rate_time_evolution}.
	b) Electron energy change rate due to the curvature term, similar to Fig.~\ref{fig:fermi_term_xy_plane}.
	c) Current density $J_z$ at $t=40\Omega_{ci}^{-1}$.
	d) Electron energy spectrum at $t=40\,\Omega_{ci}^{-1}$, inside the area shown as a black rectangle in c).
	\label{fig:thick_cs}
	}
\end{figure*}

In order to compare the inefficient acceleration
without filamentary structure formation by magnetic reconnection,
we simulated thicker current sheets that do not sufficiently
thin down to trigger streaming instabilities.
Let us demonstrate this by results obtained for a current sheet
two times thicker than the ones in which filamentation takes place.
The simulation domain is doubled along the $x-$direction,
in order to keep the same relative separation between the two current sheets as in the original run.
For the results see Fig.~\ref{fig:thick_cs}.

Fig.~\ref{fig:thick_cs}a) shows the normalized reconnection rate
for this simulation run. The maximum values are close to $0.04B_{\infty y}V_A$, i.e.,
much smaller than the values over  $0.5B_{\infty y}V_A$ from the main
simulation run reported in this paper with a thinner current sheet (Fig.~\ref{fig:rec_rate_time_evolution}a).
Because of the lower efficiency of energy conversion, reconnection
stays practically laminar (see $J_z$ in Fig.~\ref{fig:thick_cs}c), and the second current
sheet does not grow fast enough to interact with the first one during the time
period considered here. As a result, there is a longer period during which
reconnection can be considered  quasi-stationary ($t\sim(33-40)\,\Omega_{ci}^{-1}$), i.e.,
before the magnetic flux is exhausted and reconnection stops.
Thicker current sheets do not develop filamentary structures in them,
streaming instabilities are not triggered.
The resulting acceleration is very weak.
The maximum values of the curvature acceleration (see Fig.~\ref{fig:thick_cs}b)),
are, e.g., two orders of magnitude smaller than in  the case of the thinner
current sheet discussed in the paper (compare with Figs.~\ref{fig:fermi_term_xy_plane}).
The resulting electron energy spectrum near the X-line stays thermal,
departing only by a small amount from the initial energy spectrum
(Fig.~\ref{fig:thick_cs}d).
There is, therefore, no
efficient acceleration near X-lines  of laminar reconnection if no filamentary structures
develop in the guide-field (current) direction.

\section{Energy spectra in static electromagnetic fields}\label{sec:appendix2}

In order to prove that without current sheet thinning
and filamentation, in the absence of efficient
curvature acceleration, the resulting energy spectra becomes
unrealistic, we carried out another investigation.
Different from the previous simulation described in Appendix~\ref{sec:appendix},
we use the same physical parameters but
switching off at $t=10\Omega_{ci}^{-1}$
the electromagnetic field solver. The particles then continue
to move in static electromagnetic fields, without feedback to the plasma.
This corresponds to a test particle method.
We choose to stop the simulation at that time  $t=10\Omega_{ci}^{-1}$ in order to avoid the filamentation and enhanced curvature energization during the nonlinear stage of reconnection ($t\gtrsim13.5\Omega_{ci}^{-1}$).
We let the particles evolve until $t=15\Omega_{ci}^{-1}$.

\begin{figure*}[!ht]
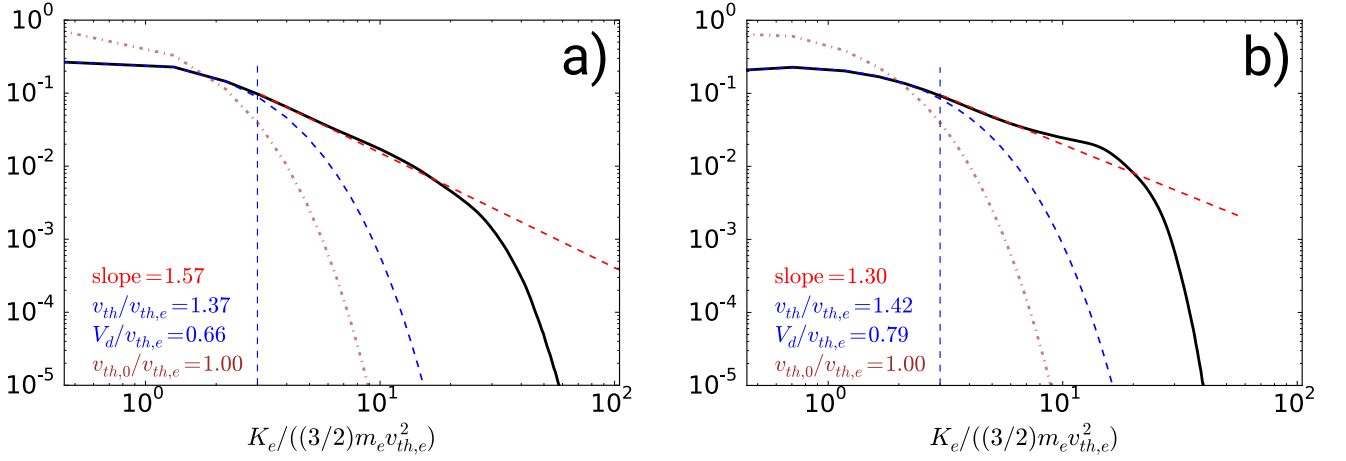

	\centering
		\includegraphics[width=0.99\linewidth]{{{fig10}}}
		\caption{Comparison of electron energy spectra close to the X-point.
	a) Two stage acceleration as in Fig.~\ref{fig:spectra_locations}.
	b) Test run with electromagnetic solver disabled.
	\label{fig:comparison_energy_spectra}}
	\end{figure*}

Following the same procedure as in Fig.~\ref{fig:spectra_locations}d),
the energy spectra are calculated in a region close to the X-point.
Fig.~\ref{fig:comparison_energy_spectra} compares the results
of this test run with the one shown before.
Fig.~\ref{fig:comparison_energy_spectra}b) shows the following
differences:
first, a heating at lower energies,
followed by a ``bump'' at higher
energies, while the electron energy spectra
drops very steeply instead of developing a power law.
The electrons attain less energy compared to the correct self-consistent consideration of the feedback.
The ``bump'' indicates a lack of particle feedback, as is commonly seen in test particles simulations
of magnetic reconnection~\citep{Kowal2012,Ripperda2017, Zhou2018}.
It is an indication of bulk acceleration of a large number of particles
to a certain energy by parallel electric fields.
This forms a beam drifting at a continuously increasing unlimited speed
due to the lack of feedback due to plasma instabilities.
The result is an artificial beam acceleration,
without the formation of power-laws.

Hence, if the curvature acceleration
is artificially suppressed, no significant power law is formed.
Instead, only heating takes place
and run-away beam formation by parallel electric fields.
Due to the lack of feedback of the particles to the electromagnetic fields,
the drift speed of the beam is continuously increasing.
Curvature-type acceleration is, therefore,
an essential element for the formation of power-laws
in a two-stage acceleration process.

\section{Validity of the guiding center approximation}\label{sec:appendix3}

	The guiding center approximation
	is only valid when the particles are fully adiabatic (fully magnetized).
	These conditions were discussed by, e.g., \cite{Northrop1963}.
	In case of a low-plasma-$\beta $ (due to the large guide field)
	the electron thermal gyro-radius in the total magnetic field,
	$\rho_{B_T}=v_{th,e}/\Omega_{ce,B_T}$,
	can be much smaller than the typical scales of the
	plasma and magnetic field variation (between $d_e$ and $d_i$).
	In our case thin current sheet with $\beta_e=0.016$,
	we initially have
	$\rho_{B_T}=d_e/11.2=d_i/112$.
	The electron gyroradius
	in the total magnetic field is just resolved by the grid cell size.
	Hence, no structures are formed at scales smaller than that length
	in our investigations.

\begin{figure*}[!ht]
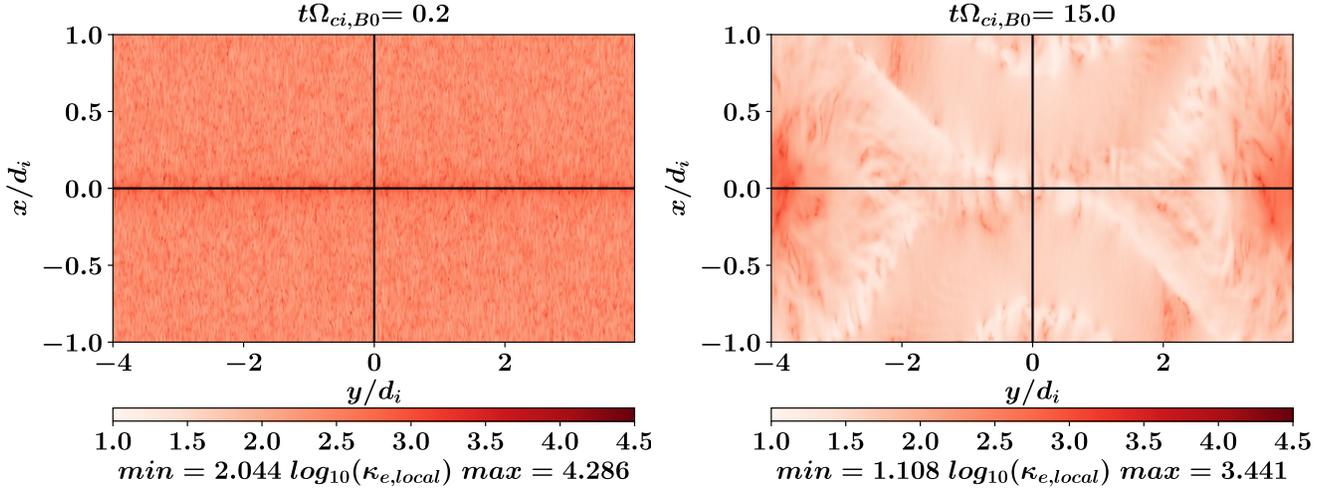
\centering
		\includegraphics[width=0.99\linewidth]{{{fig11}}}
		\caption{Contour plots of the logarithm of the $\kappa$ parameter for one slice in the reconnection plane for two different times. a) Initial $t=0.2\Omega_{ci}^{-1}$. b) Later $t=15.0\Omega_{ci}^{-1}$.\label{fig:kappa}}
	\end{figure*}

	The adiabaticity condition is fulfilled all the way in the course of
	electron acceleration since their thermal gyroradii increase only
	by small amounts.
	Let us demonstrate this by means of the calculation of the $\kappa$ parameter, introduced in
	~\citet{Buchner1989} for antiparallel reconnection and generalized in
	~\citet{Buchner1991,Buchner1991a} for guide-field reconnection geometries:
	\begin{align}\label{eq:kappa}
		\kappa=\sqrt{\Omega_{\rm min}/\omega_{\rm max}}={\rm min}\left(\sqrt{R_B/\rho_{e,{\rm eff}}}\right),
	\end{align}
	where $\Omega_{\rm min}$ is the maximum
	gyro-frequency in the minimum magnetic field strength region,
	$R_B=1/|\hat{b}\cdot\vec{\nabla} \hat{b}|$ the curvature radius of the magnetic field lines,
	$\hat{b}=\vec{B}/B$ is the unit vector in the direction of the local magnetic field,
	$\rho_{e,{\rm eff}}=v_{th,e,{\rm eff}}/\Omega_{ce}=(\sqrt{k_B T_{e,{\rm eff}}/m_e})(m_e/(eB))$
	is the electron Larmor radius in the total local magnetic field $B$,
	and $T_{e,{\rm eff}}=(1/3)(T_{e,xx} + T_{e,yy} + T_{e,zz})$  is the trace of the temperature tensor.
	The minimum is taken along the magnetic fields lines.
	$\kappa<1$ corresponds to meandering orbits,
	while $1\lesssim \kappa\lesssim 2.5$ to weakly magnetized but chaotic electrons.
	Finally $\kappa>2.5$ indicates fully magnetized electrons.
	We measured a global minimum from, initially $\kappa\sim 109$,
	to $\kappa\sim 13$ at $t=15\Omega_{ci}^{-1}$ (see Fig.~\ref{fig:kappa}).
	This is much larger than the value of $2.5$ indicating fully-magnetized electrons.
	Note that two contradicting tendencies influence $\kappa$
	during the current sheet evolution:
	the formation of small scale structures decreases
	the magnetic field curvature (small $R_B$), while the electron
	gyroradius $\rho_{e,{\rm eff}}$ increases due to the electron heating. The first effect dominates,
	resulting in an overall decrease of $\kappa$.
	Its minimum value decreases over one order of magnitude during the
	current sheet evolution.

	Those are the values calculated for thermal particles.
	For energetic particles, mainly the parallel
	velocity is enhanced, while the perpendicular
	velocity components do not vary significantly
	(see, e.g., Fig.~\ref{fig:trajectories}c).
	Therefore, their gyroradii
	do not change significantly as well
	until pitch angle scattering takes place, which
	is weak in the strong guide field limit.

	These large values of $\kappa$ indicate the remaining magnetization
	of the electrons (adiabaticity) and the validity of the guiding-center approximation throughout the acceleration process.

\end{document}